\documentclass[12pt]{article}

\usepackage{graphicx,amssymb,url,mathrsfs, amsmath}
\usepackage{eucal}
\usepackage{wrapfig}
\usepackage{boxedminipage}
\usepackage{setspace}
\usepackage{subfigure}
\usepackage[dvipdfm]{hyperref}

\def\l{\left}
\def\r{\right}
\def\t{\mathbf{t}}

\newcommand{\arctanh}[1]{\text{arctanh}}

\begin{document}

\begin{titlepage}

\begin{flushright}
{\tt hep-th/...}
\tt {FileName:....tex} \\
{\tt \today}

\end{flushright}
\vspace{0.5in}

\begin{center}
{\large \bf Holographic Jets in an Expanding Plasma}\\
\vspace{10mm}
Alexander Stoffers and Ismail Zahed\\
\vspace{5mm}
{\it \small Department of Physics and Astronomy, Stony Brook University, Stony Brook, NY 11794, USA}\\
          \vspace{10mm}
{\tt \today}
\end{center}
\begin{abstract}
We use the holographic principle to study  quark jets using trailing strings in an expanding plasma that asymptotes Bjorken hydrodynamics.
We make use of the fact that the trailing string is the locus of the light delay in bulk to obtain the explicit form for quark jets in the expanding 
plasma.  From the trailing string solution we calculate the drag coefficient of a heavy quark in the strongly coupled expanding plasma.
The energy scaling of the maximum penetration length of an ultrarelativistic light quark jet using light rays in bulk is estimated.
\end{abstract}
\end{titlepage}

\renewcommand{\thefootnote}{\arabic{footnote}}
\setcounter{footnote}{0}



\section{Introduction}

The strongly coupled quark-gluon plasma at RHIC/LHC is believed to be described appropriately by hydrodynamics at proper times $ 0.5\,\, {\rm fm }/ c \leq \tau$. One solution to  hydrodynamics describing a relativistic system with infinite extent in the perpendicular direction that is expanding longitudinally was given by Bjorken \cite{Bjorken:1982qr}.  An important probe for the strongly coupled quark-gluon plasma is the quenching of quark and gluon jets~\cite{Bjorken:1982qr}. 

A useful tool to study real time dynamics of a particle within a strongly coupled medium is the holographic approach. 
Modeling jet quenching of ultrarelativistic quarks and gluons in holography by light rays falling towards a static black hole in holography was initially 
suggested in~\cite{Sin:2004yx,SIN} and more recently in~\cite{Hatta:2007cs, Hatta:2008tx}. Jet quenching of (non-)relativistic quarks and gluons
has been described by trailing strings~\cite{Gubser:2006bz, Herzog:2006gh} and trailing and falling strings~\cite{Gubser:2008as,  Chesler:2008wd, Chesler:2008uy}; see also \cite{Ficnar:2012nu,Ficnar:2012yu}. For another approach to jet quenching within holography see \cite{Arnold:2010ir, Arnold:2011qi}. The dynamics of quark jets at early stages of their evolution was studied in \cite{Guijosa:2011hf}. Falling strings are commensurate with falling light rays in the ultrarelativistic limit bridging the gap between the two approaches.

The trailing string solution reveals momentum and energy dispersion relations for a heavy quark being dragged through a static, hot medium at a constant velocity.
For a thermal medium described by a static black hole, the drag coefficient following from the longitudinal tension of the dragged string is in 
agreement with expectations from Brownian motion~\cite{CasalderreySolana:2006rq} thanks to the Einstein relation.

The trailing string configuration can be thought of as the locus of the time delay of a light signal propagating along the jet direction in bulk.
Indeed, static quarks and gluons are threaded by a vertical color string in the holographic direction. As they are made to move longitudinally in
space, the vertical string follows causally with each point at a given holographic depth traveling with the speed of light corrected by the refraction
index of the holographic medium. As a result, the string traces the locus of time delays~\cite{SIN}. 

This particular interpretation was made clear  in~\cite{Hatta:2007cs, Hatta:2008tx}. The surface of stationary phase of the solution to 'electromagnetic' waves (more precisely $\mathcal{R}$-current)  propagating in bulk is identical to the trailing string solution.  We make use of this particular interpretation to construct trailing string solutions in a time dependent black hole background, which is dual to a time dependent expanding thermal plasma that asymptotes Bjorken hydrodynamics. 
Our explicit string construction supplements and confirms the non-equilibrium analysis of quark dragging in~\cite{Kim:2007ut}.

In section 2 we show how to obtain the trailing string solutions by following light rays propagating in bulk. Instead of using the full Maxwell description of 'electromagnetic' waves as in \cite{Hatta:2007cs, Hatta:2008tx}, it suffices to study null geodesics in bulk 
as originally suggested in~\cite{Sin:2004yx} in order to reproduce the solution to the string equations of motion. This 'eikonal approximation' holds for trailing strings in a static as well as expanding hot medium. 
In section 3, we then construct the trailing string solution in a time dependent, thermal background, which describes an expanding, hot medium, that asymptotically agrees with Bjorken's solution for large proper times. Energy and momentum dispersion relations, as well as the drag coefficient are obtained. Another trailing string solution to this setup was worked out in \cite{Giecold:2009wi}, which we briefly review in the Appendix.
The energy scaling of the stopping distance for a quark jet in holography was studied in \cite{Gubser:2008as, Chesler:2008uy, Arnold:2010ir, Arnold:2011qi}. In section 4 we approximate the stopping distance for an ultrarelativistic light quark jet by analyzing a classical particle falling in bulk on a null trajectory.  Our conclusions are given in section 5.

\section{Strings made out of light}
\begin{figure}[t]
  \begin{center}
  \includegraphics[width=12cm]{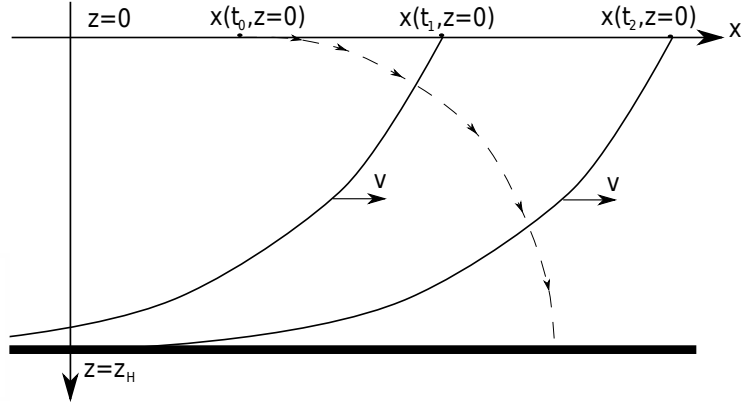}
  \caption{Light ray (dashed line) traveling from the boundary though the bulk. The trailing string (solid line) is understood as the envelope of light rays emitted from the boundary. See text.}
  \label{figure}
  \end{center}
\end{figure}

Motivated by the earlier idea on radiation in holography \cite{Sin:2004yx} and the recent 
wave analysis in~\cite{Hatta:2007cs, Hatta:2008tx}, we propose a simple way to construct the trailing string solution by looking at null geodesics in bulk. We will now show that the solution to the string equations of motion, which is a string dual to a heavy quark moving on the boundary  with constant velocity $v(t)=v$, is given by the envelope of the endpoints of light rays emitted from the quark at the boundary at an earlier time $t-t_d$, where $t_d=t_d(z)$ is the time (delay) it takes light to travel along a null geodesic from the boundary ($z=0$) to some finite $z$ in bulk. The position of the string extending into the bulk is equal to the position of an observer who has first detected the light signal emitted from the boundary at an earlier time. We illustrate this in Fig. \ref{figure}.

\subsection{Static medium}

We parametrize the position of the light ray by $\l(t(\lambda), x (\lambda), z(\lambda)\r)$ with $\lambda$ an affine parameter. The AdS$_5$-Schwarzschild metric characterizing a static medium at finite temperature reads
\begin{eqnarray}
ds^2 = \frac{R^2}{z^2} \l(- f(z) dt^2 + dx^2 + \frac{1}{f(z)} dz^2 \r) \ , \label{Adsmetric}
\end{eqnarray}
with $R$ the AdS radius, $f(z) = 1-\l(\frac{z}{z_H}\r)^4$ and $z_H$ the position of the black hole horizon. The motion along a null geodesic carries the two integrals of motion
\begin{eqnarray} \nonumber
\Pi_t &=& \frac{R^2}{z^2} f(z) \  \partial_\lambda t   \ , \\ \label{constants}
\Pi_x &=& \frac{R^2}{z^2} \partial_\lambda
 x \ .
\end{eqnarray}
 We can rewrite (\ref{constants}) as 
\begin{eqnarray} \label{xdot}
\dot {x} (t,z)\equiv \partial_t x(t,z)  =\frac{\partial_\lambda x}{\partial_\lambda t}=f(z)\frac{\Pi_x}{\Pi_t}=f(z) \ , 
\end{eqnarray} 
since the velocity of light at the boundary $z=0$ is equal to one.\\
The light ray follows the null geodesic 
\begin{eqnarray}
f(z) \l(\partial_\lambda t\r)^2 = \l(\partial_\lambda x\r)^2 + \frac{1}{f(z)} \l(\partial_\lambda z\r)^2 \ , \label{nullgeodesic}
\end{eqnarray}
resulting in $\dot{z}= \pm \l(\frac{z}{z_H}\r)^2 f(z)$.\\
At a given time $t$, we interpret the time delay, $t_d(t, z)$, as 
\begin{eqnarray} 
\partial_z t_d(t, z) \equiv \frac{\dot{x}(z=0)-\dot{x}(z)}{\dot{z}}=\frac{1-\dot{x}(z)}{\dot{z}}= \pm \frac{z^2}{z_H^2 f} \ . \label{timedelaynonexpanding}
\end{eqnarray}
If we assume a constant velocity $v$ for the quark on the boundary, each point of the wave front in bulk will move with the same velocity and $t_d(t,z) = t_d(z)$. The solution to the string $x_s (t,z)$ can now be written as
\begin{eqnarray}
x_s (t,z) = x_i + v (t-t_d) 
\end{eqnarray} 
to regain the solution to the string equations of motion \cite{Gubser:2006bz, Herzog:2006gh} with $x_i$ a constant.

\subsection{Expanding medium}
The above argument for the reconstruction of the trailing string as the locus of light delay in bulk, may be used to derive anew the
trailing string in a time dependent background composed of a falling black hole in AdS$_5$.\\
The dual geometry to a perfect fluid valid for large times with asymptotic Bjorken hydrodynamics was worked out in \cite{Janik:2005zt} and the metric in Fefferman-Graham coordinates reads
\begin{eqnarray}
ds^2= \frac{R^2}{z^2} \l[\frac{- \l( 1-\nu^4 \r)^2}{\l(1+\nu^4\r)} d\tau^2 + \l(1+\nu^4 \r) \l(\tau^2 dy^2 + dx_\perp ^2 \r) + dz^2 \r] , \label{metricJP}
\end{eqnarray} 
with the AdS radius $R$, proper time $\tau=\sqrt{t^2-\l(x^3\r)^2}$, rapidity $y=\arctanh{} \frac{x^3}{t}$ and transverse coordinates $x_\perp = \l( x^1,x^2 \r)$. The scaling variable $\nu=\frac{z}{\l(\tau/\tau_0\r)^{\frac{1}{3}}} \epsilon^{\frac 1{4}}$, with energy density $\epsilon=\frac 1{4} \l(\pi T_0 \r)^4$ of the medium and initial temperature $T_0$ at an initial proper time $\tau_0$, reveals an infalling black hole, since the horizon $z_H = \frac{\sqrt{2}}{\pi T_0}\l(\frac{\tau}{\tau_0}\r)^{\frac{1}{3}}$ is moving away from the boundary. We note that a naive interpretation of a time dependent temperature gives $T(t)\propto z_H^{-1} \propto t^{-1/3}$.  A change of variables
\begin{eqnarray}
\l( \pi^2 T_0^2 \r) w^2 = \frac{2\nu^2}{1+\nu^4} \ ,  \\ \nonumber \label{newcoordinates}
\frac{\t}{\t_0}=\frac{3}{2} \l(\tau/\tau_0 \r)^{\frac2{3}} =\frac{3}{2} \l(t/t_0 \r)^{\frac2{3}} \ ,
\end{eqnarray}
and neglecting terms $\mathcal{O}(\frac{1}{\tau})$ yields \cite{Kim:2007ut}
\begin{eqnarray}
ds^2 = \frac{R^2}{w^2} \l[- f(w) d\t^2 + \frac{4}{9} \t^2 dy^2 + \frac 3{2}\frac{\t_0}{\t} dx_\perp^2 + \frac{1}{f(w)}dw^2 \r]\label{metric} \ ,
\end{eqnarray}
with $f(w)=1-(\pi^4 T_0^4) w^4 \equiv 1-\l(\frac{w}{w_H}\r)^4$. This geometry bears a static black hole, i.e. the horizon $w_H$ is $\t$-independent, with the time dependence, $\t=\t(\tau)$, absorbed into the spacial parts of the metric. \\

In the background described by the metric in eq. (\ref{metric}), the null geodesic at the boundary gives $\partial_\t x (\t, w=0) \propto \sqrt{\frac{2 \t}{3 \t_0}}$ and leads us to introduce the time delay  as
\begin{eqnarray} \label{timedelay2}
\t_d (\t, x) = x(\t, w=0) - x =  \t_0 \sqrt{\frac{2}{3}}\l(\frac{2 \t}{3 \t_0}\r)^{\frac{3}{2}} - x  \ .
\end{eqnarray} 
We define $\hat{\t}_d=\sqrt{\frac{3\t_0}{2\t}}\t_d$ and neglecting contributions $\mathcal{O}\l(\frac{\hat{\t}_d}{\t}\r)$, which are suppressed at late proper times, the metric (\ref{metric}) at large times $\t$ assumes the form
\begin{eqnarray}
ds^2 \simeq \frac{R^2}{w^2} \l[\l(1- f(w)\r) d\t^2 - 2 d\hat{\t}_d \ d\t + d\hat{\t}_d^2 + \frac{1}{f(w)}dw^2 \r] \ .
\end{eqnarray}
For a light ray $(\t(\lambda), \hat{\t}_d(\lambda), w(\lambda))$ in this background, the two integrals of motion and the null geodesic equation read
\begin{eqnarray} \label{integral1}
\Pi_{\t} &=& \frac{R^2}{w^2} \l((1-f)\partial_\lambda \t - \partial_\lambda \hat{\t}_d \r) \ , \\ \label{integral2}
\Pi_{\hat{\t}_d} &=& \frac{R^2}{w^2} \l(\partial_\lambda \hat{\t}_d - \partial_\lambda \t \r) \ , \\ \label{nullgeodesic2}
0&=&(1-f) + \l(\partial_\t \hat{\t}_d\r)^2 - 2 \partial_\t \hat{\t}_d + \frac{\l(\partial_\t w \r)^2}{f} \ .
\end{eqnarray} 
The ratio of $\Pi_{\t}$ and $\Pi_{\hat{\t}_d}$ is fixed on the boundary
\begin{eqnarray}
\frac{\Pi_{\t}}{\Pi_{\hat{\t}_d}}\Big|_{w=0} = \frac{1-f - \partial_\t \hat{\t}_d}{\partial_\t \hat{\t}_d-1}|_{w=0} =\frac{\partial_\t \hat{\t}_d}{1+\partial_\t \hat{\t}_d}=0 \ ,
\end{eqnarray}
since the time delay $\t_d$, (\ref{timedelay2}), and $\partial_\t \t_d$ vanish on the boundary by construction. Since $\frac{\Pi_{\t}}{\Pi_{\hat{\t}_d}}$ is constant everywhere, we obtain $\partial_\t \hat{\t}_d= 1-f$ and eq. (\ref{nullgeodesic2}) results in
\begin{eqnarray}
\l( \partial_\t w\r)^2 = \pm f^2 (1-f) \ .
\end{eqnarray}
We can now solve for the time delay:
\begin{eqnarray} 
\partial_w \hat{\t}_d = \frac{\partial_\t \hat{\t}_d}{\partial_\t w} =  \pm \frac{w^2}{w_H^2 f}  \ .
\end{eqnarray}
We recognize the same equation as for the bulk part of the trailing string with a constant velocity at the boundary, (\ref{eomxi1}). The time delay, (\ref{timedelay2}), allows us to construct the following solution
\begin{eqnarray}
x_s(\t,w) = x_i + v \l(\t_0 \sqrt{\frac{2}{3}}\l(\frac{2 \t}{3 \t_0}\r)^{\frac{3}{2}}- \sqrt{\frac{2\t}{3\t_0}}\hat{\t}_d\r) \ ,
\end{eqnarray}
which matches the solution to the string equations of motion, compare (\ref{constantvelocitysolution1}). 

\section{Trailing strings in an expanding plasma} 

We now focus on the dynamics in the $x_\perp$-direction. In the static gauge, the Nambu-Goto action governs the dynamics of a string profile parametrized by $x_\perp \equiv x (\t,w)$ in the background (\ref{metric}) and reads with the Ansatz $x(w,\t)=x_i +  v \t_0 \sqrt{\frac{2}{3}} \l( \frac{2 \t}{3 \t_0} \r)^{\frac 3{2}} + \sqrt{\frac{2 \t}{3 \t_0}} \xi(w)$ 
\begin{eqnarray}
S &=& -\frac{1}{2\pi \alpha'} \int d\t dw \sqrt{-g} \\
&=&- \frac{1}{2\pi \alpha'} \int d\t dw \frac{R^2}{w^2} \sqrt{1-\l(\frac{3 \t_0}{2\t}\r) \frac{1}{f(w)} \l(\partial_\t x\r)^2 + \l(\frac{3 \t_0 }{2 \t}\r) f(w) \l(\partial_w x\r)^2} \\ 
&\simeq & - \frac{1}{2 \pi \alpha'} \int d\t dw \frac{R^2}{w^2}\sqrt{1- \frac{v^2}{f(w)} + f(w) \xi'(w)^2} \ , \label{actionconstantvelocity}
\end{eqnarray}
up to corrections of $\mathcal{O}(\frac{\xi}{\t})$. Gravity will pull the string towards the horizon increasing $\xi(w)$ as $w \rightarrow w_H$. Corrections of the order $\mathcal{O}(\frac{\xi}{\t})$ will show up in the infrared behavior of the solution.
In (\ref{actionconstantvelocity}), we recognize the same form that leads to the trailing string solution \cite{Gubser:2006bz, Herzog:2006gh} 
modulo a transformation $w=R^2/u$. A solution to the string profile $\xi(w)$ is easily worked out; \cite{Gubser:2006bz, Herzog:2006gh}. The equation of motion for $\xi(w)$ reads
\begin{eqnarray}
\frac{R^4}{w^4} \frac{f(w)}{\sqrt{-g}} \partial_w \xi(w) &= &c(\t) \\ \label{eomseparable1}
\Leftrightarrow \partial_w \xi(w) &=& \pm  \frac{c}{f} \frac{w^2}{R^2} \l(\frac{f-v^2}{f-c^2 \frac{w^4}{R^4}} \r)^{\frac{1}{2}} \ ,
\end{eqnarray}
with $c$ an integration constant. Demanding the energy and momentum flow, (\ref{energymomentumloss}), and, thus, $\partial_w \xi$ to be positive and real, fixes $c$ to be 
\begin{eqnarray}
c = \frac{R^2}{w_c^2} v= \frac{R^2}{w_H^2} \sqrt{\frac{v^2}{1-v^2}} \ , 
\end{eqnarray}
where $w_c^4= w_H^4 \l(1-v^2 \r)$ is the critical 'radius' at which both numerator and denominator in the bracket in (\ref{eomseparable1}) change sign. The equation of motion (\ref{eomseparable1}) reads
\begin{eqnarray}
\partial_w \xi(w) = \frac{v}{w_H^2} \frac{w^2}{f}  \label{eomxi1}
\end{eqnarray}
and yields the solution to the string
\begin{eqnarray}
x(w,\t)=x_i + v \t_0 \sqrt{\frac{2}{3}} \l( \frac{2 \t}{3 \t_0} \r)^{\frac 3{2}} + \sqrt{\frac{2 \t}{3 \t_0}} \frac{v w_H}{2} \l( \arctanh{}\frac{w}{w_H} - \arctan{} \frac{w}{w_H}\r) \ . \label{constantvelocitysolution1}
\end{eqnarray}
The on-shell action is proportional to $S \propto \int dw d\t \frac{R^2}{w^2} \sqrt{1-v^2}$. At the boundary, $x(t)\propto t$, representing a heavy quark at constant velocity.

\section{Energy-loss and drag coefficient}
Integrating the canonical energy and momentum densities $\pi^0_t, \pi^0_x$ \cite{Gubser:2006bz, Herzog:2006gh}, the total energy and momentum of the string in the $(\t,w)$-coordinate system are given by
\begin{eqnarray}\label{energy} 
E &=& - \int dw \ \pi^0_t = \frac{\sqrt{\lambda}}{2\pi} \int^{w_H}_{w_\epsilon} \frac{dw}{\sqrt{-g}} \frac{R^2}{w^4} \l(1+ \frac{3 \t_0}{2 \t} f(w) \l(\partial_w x\r)^2 \r) \ , \\ \label{momentum}
p &=& \int dw \ \pi^0_x = \frac{\sqrt{\lambda}}{2\pi} \int^{w_H}_{w_\epsilon} \frac{dw}{\sqrt{-g}} \frac{R^2}{w^4} \frac{3 \t_0}{2 \t} \partial_\t x \l(1 + \frac{w^4}{w_H^4 f(w)} \r) \ ,
\end{eqnarray}
with $\sqrt{\lambda}=\frac{R^2}{\alpha'}$
The static solution $x(\t,w)=x_i=const.$ defines the static thermal mass $M(T_0)$ as
\begin{eqnarray}
M(T_0)=- \int dw \ \pi^0_t= \frac{\sqrt{\lambda}}{2\pi} \l(\frac{1}{w_\epsilon} - \pi T_0\r) \equiv m - \Delta m_0 \ .
\end{eqnarray}
The static solution, as well as the non-static one above, has Dirichlet boundary conditions, implying that the endpoint of the string does not vary with $w$. This requires $w_\epsilon = const. \Leftrightarrow z_\epsilon \propto t^{1/3}$. As the black hole moves away from the boundary ($z_H=z_H(t)$) in the $(t,z)$-frame, the endpoint of the string has to reflect this dynamic behavior.
At large times the medium is described by (Bjorken) hydrodynamics satisfying the assumption of near local equilibrium. Static properties of the quark are insensitive to the changes in the medium as the thermal mass correction, $\Delta m_0=\Delta m_0(T_0)$, is constant in time. Interpreting $\Delta m_0$ as a screening mass, we would expect it to be proportional to $T(t)$. 
The associated energy and momentum flow of the string are
\begin{eqnarray} \nonumber
\partial_\t E &=& \pi^1_t = \frac{\sqrt{\lambda}}{2\pi \sqrt{-g}} \frac{R^2}{w^4} \frac{3 \t_0}{2\t} f(w) \partial_w x \ \partial_\t x \ , \\ 
\partial_\t p &=& -\pi^1_x = \frac{\sqrt{\lambda}}{2\pi \sqrt{-g}} \frac{R^2}{w^4} \frac{3 \t_0}{2\t} f(w)\partial_w x \ . \label{energymomentumloss}
\end{eqnarray}
Ignoring corrections $\mathcal{O}(\frac{\xi}{\t})$, the energy and momentum dispersion relations and flows are
\begin{eqnarray}\label{energydispersionrelationconstant} 
E &=& \gamma M(T_0) + \frac{1}{v} \l(\partial_\t E\r) \Delta \xi \ , \\ \label{energyflow} 
\partial_\t E &=&  \frac{\sqrt{\lambda}}{2 \pi} \gamma\frac{v^2}{w_H^2} \ , \\ \label{energydispersionrelationconstant2}
p &=& \sqrt{\frac{3 \t_0}{2\t}} v \gamma M(T_0)  + \frac{1}{v} \l(\partial_\t p\r) \Delta \xi \ ,  \\  
\partial_\t p &=& \frac{\sqrt{\lambda}}{2 \pi} \sqrt{\frac{3 \t_0}{2\t}} \gamma \frac{v}{w_H^2} \ ,
\end{eqnarray} 
with $\gamma=\frac{1}{\sqrt{1-v^2}}$, and $\Delta \xi=\int_{w_\epsilon}^{w_{max}} dw \  \partial_w \xi$.
$w_{max}$ is the maximum depth beyond which $\mathcal{O}(\frac{\xi}{\t})$ corrections can no longer be neglected. Note that although the string endpoint falls in the $(t,z)$-frame, the energy flow down the string is independent of the radial coordinate. While the total energy of the quark is constant in the ($\t, w$)-frame, the momentum decreases with time.As the external electric force $F_{tx}=- \pi^1_x$ weakens with time, the quark loses momentum and is asymptotically static. (Note that the medium does not expand in the $x_\perp$ direction.) 

We can now evaluate the drag coefficient $\mu$. The momentum flowing away from the quark on the boundary reads
\begin{eqnarray}
\frac{dp}{dt} &=& \l(\frac{\partial p }{\partial \t}\r)\l(\frac{\partial \t }{\partial t}\r)  = \pi^1_x \l(\frac{\partial \t }{\partial t}\r)\\
&=& \frac{- \sqrt{\lambda}}{2\pi} \frac{(\pi T_0)^2}{M(T_0)} \l(\frac{\t_0}{t^{\frac{1}{3}}t_0^{\frac{2}{3}}}\r) \sqrt{\frac{3 \t_0}{2\t}} \frac{v M(T_0)}{\sqrt{1-v^2}} \ .
\end{eqnarray}
For a quark momentum $p=\sqrt{\frac{3 \t_0}{2\t}} v \gamma M(T_0)$ we obtain the drag coefficient
\begin{eqnarray}
\frac{dp}{dt} &=& - \mu p \\
\mu M (T_0)&=& \frac{\sqrt{\lambda}}{2\pi} (\pi T_0)^2 \l(\frac{\t_0}{t^{\frac{1}{3}}t_0^{\frac{2}{3}}}\r) \propto \sqrt{\lambda} T_0 \ T(t)  \propto \Delta m_0 \ T(t) \label{dragcoefficient} \ . 
\end{eqnarray}
This is analogous to the momentum loss in a static background with a constant temperature $T=T_0$, where the drag is governed by $\mu M (T_0)= \pi \Delta m_0T_0$, \cite{Gubser:2006bz, Herzog:2006gh}.  The result (\ref{dragcoefficient}) is also in agreement with the Brownian motion analysis carried in~\cite{Kim:2007ut}
for a falling black hole and using different approximations.

\section{Penetration depth for a light quark jet}

For a relativistic light quark jet traveling through an (expanding) medium, we give an estimate for the maximum penetration depth. A light quark is approximately described by a null string whose endpoint is initialized at some finite radial coordinate and falls into the black hole following a null geodesic \cite{Chesler:2008wd,Chesler:2008uy}. The approximation used is that long enough after the initialization, once the endpoints of the string have traveled from the boundary to some depth $z_*$, the quarks flying away from each other can be represented by trailing string solutions, where the string uniformly translates with a constant velocity and its endpoints follow null geodesics. The excitation on the boundary will dissolve into the medium once the endpoints of the string have reached the horizon and the energy scaling of the stopping distance is $\l(\Delta x(E)\r)_{max} \approx \l({E}/{T \sqrt{\lambda}}\r)^{{1}/{3}}/T$.\\
Following the classical approximation used in \cite{Arnold:2011qi}, we approximate the maximum stopping distance of a relativistic particle on the boundary by following light rays in bulk. We first consider the case of a static medium, whose dynamics are described by the metric in (\ref{Adsmetric}). To describe a quark on the boundary with velocity $v$, we fix the constants of motion, (\ref{xdot}),
\begin{eqnarray}
\dot{x}=\frac{dx}{dt}=\frac{\Pi_x}{\Pi_t} f =v f \ .
\end{eqnarray}
The null geodesic reads  
\begin{eqnarray}
\dot{z}=\frac{dz}{dt} = \sqrt{f\l(f-\dot{x}^2\r)} \ ,
\end{eqnarray} 
and the distance $\Delta x$ the quark travels before the string reaches the horizon is given by
\begin{eqnarray}
\Delta x = \int_0^{z_H} dz \frac{\dot{x}}{\dot{z}} = \int_0^{z_H} dz \frac{v}{\sqrt{1-v^2f}} = \int_0^{z_H} dz \frac{v}{\sqrt{\frac{1}{\gamma^2}+v^2 \l(\frac{z}{z_H}\r)^4}} \ .
\end{eqnarray}
For ultrarelativistic motion, $v\simeq 1$ and finite $\frac{1}{\gamma}$, the energy dependence of the maximum penetration depth reads
\begin{eqnarray} \label{penetrationdepth}
\l(\Delta x\r)_{max} \approx \sqrt{\gamma} z_H \approx  \frac{\sqrt{E/Q}}{T} \ , \
\end{eqnarray} 
where we have assumed a dispersion relation of the form $E=\gamma Q$. The result in (\ref{penetrationdepth}) agrees with the one found in \cite{Arnold:2011qi}, $\Delta x \simeq \left( \frac{E^2}{-q^2}\right)^{1/4}$, where $q^2$ was identified with the virtuality of the classical particle. However, the disperion relation for the string, compare (\ref{energydispersionrelationconstant}),(\ref{energydispersionrelationconstant2}), supports the identification $Q=(T)$. For very heavy quarks $Q$ can be identified with the bare quark mass. It would be interesting to compare these two results $\l(\Delta x\r)_{max} \simeq \sqrt{E/M}$ and $\l(\Delta x\r)_{max} \simeq \left(\frac{E^2}{-q^2}\right)^{1/4}$ with current experimental data on heavy quark jets at RHIC and LHC.

Similarly, at late times in an expanding background, (\ref{metric}), where the quark on the boundary moves close to the speed of light with $x(t,z=0)\propto v \t^{\frac{3}{2}}$, the maximum stopping distance is given by
\begin{eqnarray}
\l(\Delta x\r)_{max} \propto \frac{1}{\sqrt{1-v^2}} \frac{\sqrt{\t}}{T_0} \propto \frac{\sqrt{E/M(T_0)}}{T(t)} \ .
\end{eqnarray}

\section{Conclusion}

Falling black holes in the context of holographic QCD offer a new framework for studying
strongly coupled gauge theories out of equilibrium~\cite{Shuryak:2005ia}. Heavy quark diffusion
in such a context was originally discussed in~\cite{Kim:2007ut} using certain assumptions. The current analysis through the 
trailing string confirms these findings without additional assumptions.
Key to the construction of the trailing string solution in the time dependent holographic background
is the observation that the string in bulk is the locus of the light delays~\cite{SIN,Hatta:2007cs,Hatta:2008tx}.

The falling black hole in bulk asymptotes Bjorken hydrodynamics, describing a longitudinally expanding plasma
with infinite extent in the perpendicular spatial direction.  The time-delayed and approximate solution to the string equations
of motion accounts for a quark at constant velocity. Furthermore, an exact solution exists, representing 
a decelerated quark at the boundary with an asymptotically vanishing momentum (see Appendix).

Since the holographic metric can be expressed in a coordinate frame in which the horizon is fixed in proper time, the essentials of the
expanding medium are captured by a time dependent rescaling of the temperature. A falling black hole from $t_0$ to $t$ describes
an expanding and cooling medium from $T(t_0)=T_0$ to $T(t)\approx T_0/\tau^{1/3}$. Heavy quarks with initial thermal masses
$M(T_0)=m-\Delta m_0$ are dragged with a dragging coefficient $\mu\,M(T_0)\approx \Delta m_0\,T(t)$.  The latter is the analog of the dragging in a static medium with fixed temperature. Ultrarelativistic, on-shell quarks with energy $E$ penetrate a distance 
$\left( \Delta x\right)_{max} \approx \sqrt{\gamma}z_H\approx\sqrt{E/M(T_0)}/T(t)$ in the expanding plasma.

\section{Acknowledgements}
This work was supported in parts by the US-DOE grant DE-FG-88ER40388. 

\section{Appendix} \label{Appendix}

In this Appendix we show that the falling black hole background (\ref{metric}) also admits an
exact, decelerated trailing string, which exhibits identical dragging to the trailing string solution discussed in the text. This solution was worked out in \cite{Giecold:2009wi}.
The equation of motion for a solution of the form $x(\t, w) = const. + v \t + \xi(w) $ reads
\begin{eqnarray} 
\frac{R^4}{w^4} \l(\frac{3 \t_0}{2\t}\r) \frac{f(w)}{\sqrt{-g}} \partial_w \xi(w) &= &c(\t) 
\end{eqnarray}
and gives the same result as in (\ref{eomxi1})
\begin{eqnarray}
\partial_w \xi(w) = \frac{v}{w_H^2} \frac{w^2}{f} \ ,
\end{eqnarray}
and, thus, 
\begin{eqnarray}
x(\t, w) &=& x_i + v \t + \frac{v w_H}{2} \l( \arctanh{}\frac{w}{w_H} - \arctan{} \frac{w}{w_H}\r) \ . \label{separablesolution1} 
\end{eqnarray}  
$x_i$ is related to the initial position of the string endpoint. 
The on-shell action density corresponding to~(\ref{separablesolution1}) is given by $\sqrt{-g}=\frac{R^2}{w^2} \sqrt{1-v^2 \l(\frac{3 \t_0}{2 \t}\r)}$.
Since $\t \propto t^{\frac{2}{3}}$, the quark moving along the trajectory of the endpoint of the string at $z=0$ is decelerated. The string solution (\ref{separablesolution1}) describes a trailing string and every point on the string moves with the same velocity in the $x$-direction in the $(\t, w)$ frame. This is no longer true if we transform to the original variables $(t,z)$, since the black hole horizon is falling in these coordinates. The deceleration of the quark is proportional to the rate at which the temperature decreases, $T(t) \propto t^{-1/3}$.
For the solution in (\ref{separablesolution1}), the densities read
\begin{eqnarray}
\pi^0_t &=& \frac{\sqrt{\lambda}}{2 \pi} \frac{1}{\sqrt{1-v^2 \l(\frac{3 \t_0}{2\t} \r)}} \l(\frac{1}{w^2} + \frac{3\t_0}{2\t}\frac{v^2 w^2}{w_H^4 f} \r) \\
\pi^1_t &=& \frac{\sqrt{\lambda}}{2 \pi} \frac{1}{\sqrt{1-v^2 \l(\frac{3 \t_0}{2\t} \r)}} \l(\frac{3 \t_0}{2 \t}\r) \frac{v^2}{w_H^2} \\
\pi^0_x &=& \frac{\sqrt{\lambda}}{2 \pi} \frac{1}{\sqrt{1-v^2 \l(\frac{3 \t_0}{2\t} \r)}} \l(\frac{3 \t_0}{2 \t}\r) \frac{v }{w_H ^2 f}\\
\pi^1_x &=& \frac{-\sqrt{\lambda}}{2 \pi} \frac{1}{\sqrt{1-v^2 \l(\frac{3 \t_0}{2\t} \r)}} \l(\frac{3 \t_0}{2 \t}\r) \frac{v }{w_H^2} \ ,
\end{eqnarray}
resulting in the following dispersion relations using (\ref{energy}), (\ref{momentum})
\begin{eqnarray} \label{energydispersionrelationdecelerated}
E &=& \gamma M(T_0)+ \frac{1}{v} \l(\partial_\t E\r) \Delta \xi \ , \\ \label{momentumdispersionrelation}
p &=& \l(\frac{3 \t_0}{2\t}\r)v \gamma M(T_0) + \frac{1}{v} \l(\partial_\t p\r) \Delta \xi  \ .
\end{eqnarray}
For both the decelerated and the constant velocity solution, $v$ is related to the initial quark velocity. The energy dispersion relations (\ref{energydispersionrelationdecelerated}) and (\ref{energydispersionrelationconstant}) have the same structure, ensuring that both string solutions describe the same excitation on the boundary. 
Since both solutions probe the same medium, it is not surprising, that we obtain the same drag coefficient as in ($\ref{dragcoefficient}$) for a quark having momentum $p=\sqrt{\frac{3 \t_0}{2\t}} v \gamma M(T_0)$.

\end{document}